\documentclass[12pt]{iopart}
\usepackage[dvipdf,pdftex]{graphicx}
\begin{document}

\title[Reducing zero-drifts in DMI with thermostatic chamber]{Characterising and tackling thermally induced zero-drift in displacement measuring interferometry using temperature-controlled enclosure}

\author{Simon Rerucha\footnote{Author to whom any correspondence should be addressed}, Miroslava Hola, Ondrej Cip, Josef Lazar, Jindrich Oulehla and Bretislav Mikel}

\address{Institute of Scientific Instruments of the CAS (ISI), Kralovopolska 147, 612 00 Brno, Czech Republic }
\ead{res@isibrno.cz}

\begin{abstract}
Our research efforts in displacement measurement interferometry focused on long-term drifts initiated an extended experimental investigation in the interferometric assemblies of our design. We aimed to analyze, characterize and tackle the long-term measurement stability, expressed as the zero-drift, with particular attention to the thermal effects. For the experimentation, we developed a thermostatic chamber equipped with active temperature regulation, an array of sensors and control electronics. With either the finely stabilized temperature or with the thermal cycling, we can carry out a range of investigations:  verification of modified design or prototype interferometers, testing of production pieces, characterization of integrated assemblies and units in terms of the zero drift and the susceptibility to thermal effects -- the temperature sensitivity ($\delta L / \delta T$, expressed in nm.K$^{-1}$). With these experimental studies, we demonstrate the potential of the zero drift studies to contribute to the development and broader expansion of interferometric instrumentation.

Note: This is the version of the article before peer review or editing, as submitted by an author to Measurement Science and Technology. IOP Publishing Ltd is not responsible for any errors or omissions in this version of the 
manuscript or any version derived from it.

Published article: Simon Rerucha et al 2024, Meas. Sci. Technol. 35 115011\\
https://doi.org/10.1088/1361-6501/ad646b

\end{abstract}

%
\vspace{2pc}

\submitto{\MST}

%
\maketitle
%
%

\section{Introduction}

Interferometry, a cornerstone technique in modern length metrology, has revolutionized the field of dimensional measurement by offering unparalleled precision and accuracy \cite{yang2018review}. 
Employing the principles of wave interference \cite{michelson1887ontherelative}, interferometry enables the determination of distances and dimensions with extraordinary sensitivity, making it indispensable in a wide range of scientific and industrial applications. 
At the same time, interferometric methods have continually pushed the boundaries of achievable precision. 

With the resolution and precision that matches and even dives below the size of single atoms, there is a naturally long list of aspects and influences that require careful consideration to achieve successful interferometric measurement \cite{coveney2020review}. 
Many of these aspects have already been well covered in the literature and are still actively investigated. 

However, the question of the \textit{(long-term) stability of displacement reading}, typically expressed as the \textit{zero-drift} (e.g. in the ANSI/ASME B89.1.8 standard \cite{stone2014testcalibration}), has been so far either overlooked (for many good practical reasons) or simply not of interest. 
Reducing the zero drift is generally desirable for any interferometric instrumentation. 
Nonetheless, there are two specific paths of development where minimal drifts are mainly of interest: measurement in an environment with sub-optimal conditions and measurement over extended time frames. 
In these areas, the zero-drift is an obvious obstacle to the broader penetration of interferometry to metrological applications in the high-tech industry and generally beyond the boundaries of national metrology laboratories.

The extended time-frame examples typically occur in applications that require precise and stable positioning (large area scans in nanometrology, long expositions in e-beam lithography, reference object positioning in optical metrology), in applications where the interferometer(s) serve as a length reference (large volume metrology, calibrations with larger than the usual number of measurement points or repetitions) and also in fundamental research (e.g. research on gravitational waves \cite{figueroa2019ability}, the realization of the unit kilogram using Kibble balance \cite{robinson2016watt}). 

The other part seeks to deliver the precision, inherent traceability, and dynamic range of interferometric measurements operated in a well-controlled and monitored environment (primarily the temperature, air pressure, and relative humidity) outside the laboratories. 
The applications range from 100 $\%$ characterization of sensors elements \cite{rerucha2021ctrp} in production (as required by Industry 4.0) via fast turnover in-house calibration of working standards (e.g. gauge blocks, end bars) to in-situ calibration of instrumentation for highly specific tasks (such as structural inspection in nuclear power plants \cite{rerucha18temelin}).

Our work aims to investigate, characterize and possibly reduce zero-drifts, especially tge thermally induced ones, to make the unprecedented accuracy of the interferometric measurement applicable to these kinds of applications and measurement scenarios. Regarding the application parameters, our work generally targets the range between the micro- and nanoscale (broadly covered by the nanometrology community today) and the long-range measurements. We thus aim at the \textit{metre range}, i.e. lengths between $0,1\,$m and $10\,$m, with a particular focus on the lower part of the range that matches e.g. with the coordinate positioning. With precision and accuracy, we are targeting the contribution to the measurement uncertainty of $10^{-7}$ to reach a comparable level to the other fundamental limited, the influence of the refractive index of air. As the temperature-induced zero-drifts of an interferometer scale primarily with the arrangement and dimensions of the optical part and the magnitude of the temperature fluctuations rather than the measured distance, we practically aim at a $\approx10\,$nm.K${-1}$ target.

In this paper, we explore the length drifts beginning with the perspective of the entire measurement system assembly, quickly moving towards the system's heart, the interferometer itself. 
During the journey, we identify the main culprit: the omnipresent effects of thermal changes on both the geometrical and material characteristics (Section \ref{s:effects}). 
With this in mind, we design and realize a dedicated temperature-controlled enclosure as an experimental testbed to investigate the interferometric components' and assemblies' temperature sensitivity under thermal load (Section \ref{s:jebox}). Then we present four experimental studies, demonstrating how the zero drift studies can contribute to the characterization, verification and tackling the zero drift in interferometric systems (Section \ref{s:experimental}).
The paper is then summarised and concluded with future prospects in what we see as a broad potential for progress in interferometry-based dimensional metrology (Section \ref{s:final}).

\section{Thermal effects influencing interferometric measurement}
\label{s:effects}

As it is well known, temperature and its fluctuations profoundly influence dimensional measurement and impact the measurements' accuracy and reliability. 
With laser interferometry, the fluctuating temperature influences all components of any measurement assembly -- laser source (and delivery), the interferometer and optics, the (opto-) mechanics, the environment (air) and detection electronics -- in a way that might exhibit as thermally induced zero drifts \cite{rerucha2024interferometer}.

This influence primarily stems from the thermal expansion, which alters the geometry of both the measuring arrangement (not necessarily interferometric at this point) and the measurand. 
While the expansion of the measurand is typically accounted for (e.g. in gauge-block calibration \cite{byman2018high}), accounting for the dimensional changes that influence the structural loop and, consequently, the measurement loop of the measurement systems can be challenging and is often neglected. 

The most straightforward way to limit the thermal effects at the system level is to keep the outer environment stable (a remarkable approach could be found in Helsinki \cite{lassila2011design}) and let the measurement equipment stabilize well before the actual measurement. The disadvantage is that any local temperature disturbances (e.g. heat dissipation from electronics or translation mechanisms) interfere with the temperature control and induce uneven temperature distribution along the measurement assembly. Another (complementary rather than alternative) approach to mitigating thermal issues incorporates a careful design of the measuring system focused on a well-stabilized metrology loop so that thermal creeps have a limited influence on the measurement arrangement. Typical examples could be using low-expansion materials for the metrology frame \cite{manske2012recent}, using differential interferometers \cite{yacoot2000use,rerucha2021compact} to shorten the loop and, generally, sticking to good practice for precision engineering \cite{yague2021scalability}.

With interferometer(s) involved, the temperature also has a massive influence on the refractive index of air, as summarized in Table \ref{t:ria}. 
Unless the entire measurement system is put into vacuum \cite{weichert2018vacuumsetup}, the measurement accuracy is hampered by $\approx 10^{-6}\,\cdot\,$K$^{-1}$ \cite{schodel2021new}.
This influence is usually mitigated using a suitable empirical equation (i.e. indirect refractometry \cite{ciddor1996refractive}). Nonetheless, the residual issues with a non-uniform temperature distribution of unknown magnitude could remain (for the comparison, the climate chambers used for such purposes usually declare the temperature homogeneity somewhere at $0.1\,^{\circ}$C). For practical applications, it is, however, essential to consider the degree of precision needed for measurement of the other parameters of the air, namely the atmospheric pressure, to reduce both absolute errors and potential drifts (see the fourth column in Table \ref{t:ria} that states the relative error contribution stemming from a 1\% measurement error).

A further element of interest is the susceptibility to the thermal effects of the interferometer itself, particularly the interferometer's optics \cite{rerucha2024interferometer}.
With fluctuating temperatures, the geometrical dimensions change, and so does the refractive index of the optical material the interferometer's optics is manufactured of.  
The variation of the refractive index with temperature is referred to as the \textit{thermo-optic coefficient} \cite{ghosh1997sellmeier, palik1997thermo}.
Table \ref{t:properties} presents selected optical materials and corresponding values of the coefficient of the thermal expansion (CTE) and the thermo-optic coefficient (dn/dT \cite{palik1997thermo}).
The values indicate that these influences might have a magnitude comparable to that of the thermal expansion and the fluctuations of the refractive index of air; nonetheless, the magnitudes are not comparable directly, as the refraction influence scales with the dead-path in the air, the thermal drift of the optical system scales with its dimensions.   

To minimize the thermal drifts, it is crucial to design the optical arrangement of the interferometer in a way the parts of the interferometer's arm within the interferometer's optics are of equal optical length\cite{steinmetz1990sub,rerucha2024interferometer} and ideally share the path to a maximal extent. 
Despite these decades-old findings, it seems many state-of-the-art interferometric designs tend to neglect this aspect; examples to be found, e.g. in \cite{pisani2012comparison,rerucha2021compact}. Also, to our best knowledge, the thermal drifts were not thoroughly experimentally characterized in detail at the time of the original investigation.  

\begin{table}
\caption{The thermal properties of interest for selected optical materials used for construction of interferometers (at $633\,$nm and $20^{\circ}C$; values vary by source/manufacturer)}
\begin{indented}
\lineup
\item[]\begin{tabular}{l r r l}
\br                              
            & CTE & dn/dT &  \cr 
Material    & ($10^{-6}/$K) & ($10^{-6}/$K) & Notes \cr 
\mr
Schott N-BK7  & 7,1 & 1,29 & \cr
Schott N-SF14 & 9,41 & -1,46 & \cr
Fused silica  & 0,55 & 12,9  &  \cr
Crystal quartz & 7,5 & -5,5  & perpendicular to opt. axis\cr
               & 13,5 &  -6,5 & parallel to opt. axis\cr
\mr
Refractive index of air & -- & 0,95 &\cr
\br
\label{t:properties}
\end{tabular}
\end{indented}
\end{table}

\begin{table}

\caption{The dependence (relative change) of the refractive index of the air on the environment (temperature, humidity, and pressure) and the environmental stability (at $633\,$nm and nominal values; based on \cite{schodel2021new})}

\begin{indented}
\lineup
\item[]\begin{tabular}{l c r r}
\br                              
                    & Value at standard & Refractive index  & R. I. change per $1\%$ \cr 
Influence parameter & conditions  & sensitivity coefficient  & value change   \cr 
\mr
Temperature  & $20\,^{\circ}$C  & $-0,92\times10^{-6} $K$^{-1}$ & $-0.19\times10^{-6}$  \cr
Pressure     & $101 325\,$Pa & $2,7\times10^{-9}$Pa$^{-1}$ & $2,72\times10^{-6}$  \cr
Relative humidity & $50\,$\%  & $-8,7\times10^{-9}$(\% RH))$^{-1}$ & $<0,01\times10^{-6}$ \cr
CO$_2$ content & $400\,$ppm  & $1,4\times10^{-10}$(ppm))$^{-1}$ & $<0,01\times10^{-6}$ \cr
\br
\label{t:ria}
\end{tabular}
\end{indented}
\end{table}

Finally, the optical / opto-mechanical assembly technique can also render the interferometer sensitive to temperature changes.
An interferometer assembled from individual components (rather than cemented or contacted together \cite{rerucha2021compact}) could be potentially susceptible to thermal drifts due to uneven influence of local temperature or pressure change or some misalignments that would, e.g. not exhibit on the fringe contrast directly.  
This susceptibility will differ depending on the particular arrangement and vary piece by piece due to the manufacturing process. 

In a real-world experiment or application, all the temperature-induced effects sum up, and the resulting superposition contributes to the zero drift. Due to non-ideal aspects of reality, every compensation mechanism has its limits. Any interferometric measurement will inevitably suffer a certain amount of residual thermal drift and exhibit some level of (dominantly thermally-induced) zero-drifts. These residual drifts become more pronounced in the scenarios mentioned above with the measurement over a more extended period of time and in an environment with sub-optimal thermal conditions. 

The challenge we accepted was to specify and investigate a methodology allowing us to experimentally test the components of interferometric systems in an environment with stable and controlled temperatures. 
This approach is intended to characterize both the systematic and residual thermal drifts and quantify them as a measure called the \textit{temperature sensitivity}, expressed as a displacement depending on the temperature in $nm$.K$^{-1}$. 

The quantitative knowledge of this value is directly applicable as an input to the uncertainty analysis when the interferometric system is integrated into a measurement instrument. 
It is also helpful for the iterative development or improvement of either a single component (subsystem) or gradually extended assemblies (e.g. from standalone interferometer's optics to integrated interferometric assembly including fixtures), where it is possible to evaluate the gradual increase (desirably minimal) of the temperature sensitivity. In this case, even a quantitative estimation of the thermal sensitivity could aid the process, allowing for side-by-side comparison.
 Last but not least, the established (and experimentally verified) knowledge can serve as a reference figure for the verification and testing in the manufacturing and production of the interferometers. 

\section{Temperature-controlled experimental enclosure}
\label{s:jebox}

Typical instrumentation used to achieve temperature control in an air environment is some kind of climate chamber. Besides the costs, the major disadvantage is the inhomogeneity of the temperature inside the box in conjunction with forced air circulation, which is generally not favourable for interferometric measurements. 

To overcome these limitations and allow for the characterization of the thermal sensitivity of interferometric systems, we have designed a temperature-controlled experimental enclosure to achieve thermal stability better than $50\,$mK. 
This (intermittent) goal was chosen so that the relative accuracy of displacement measurement, burdened with the thermal expansion and fluctuations of the air refractive index, would be at a similar level as the other significant source of drifts: the frequency source ($2 \cdot 10^{-8}$ with a frequency-stabilized He-Ne laser). In the following sections, we briefly present the design and construction of the enclosure, the achievable thermal condition within the enclosure, and the methodology we used for the experimentation.

\subsection{Enclosure design}

\begin{figure}[htbp]
	\begin{center}
		\includegraphics[width=\textwidth]{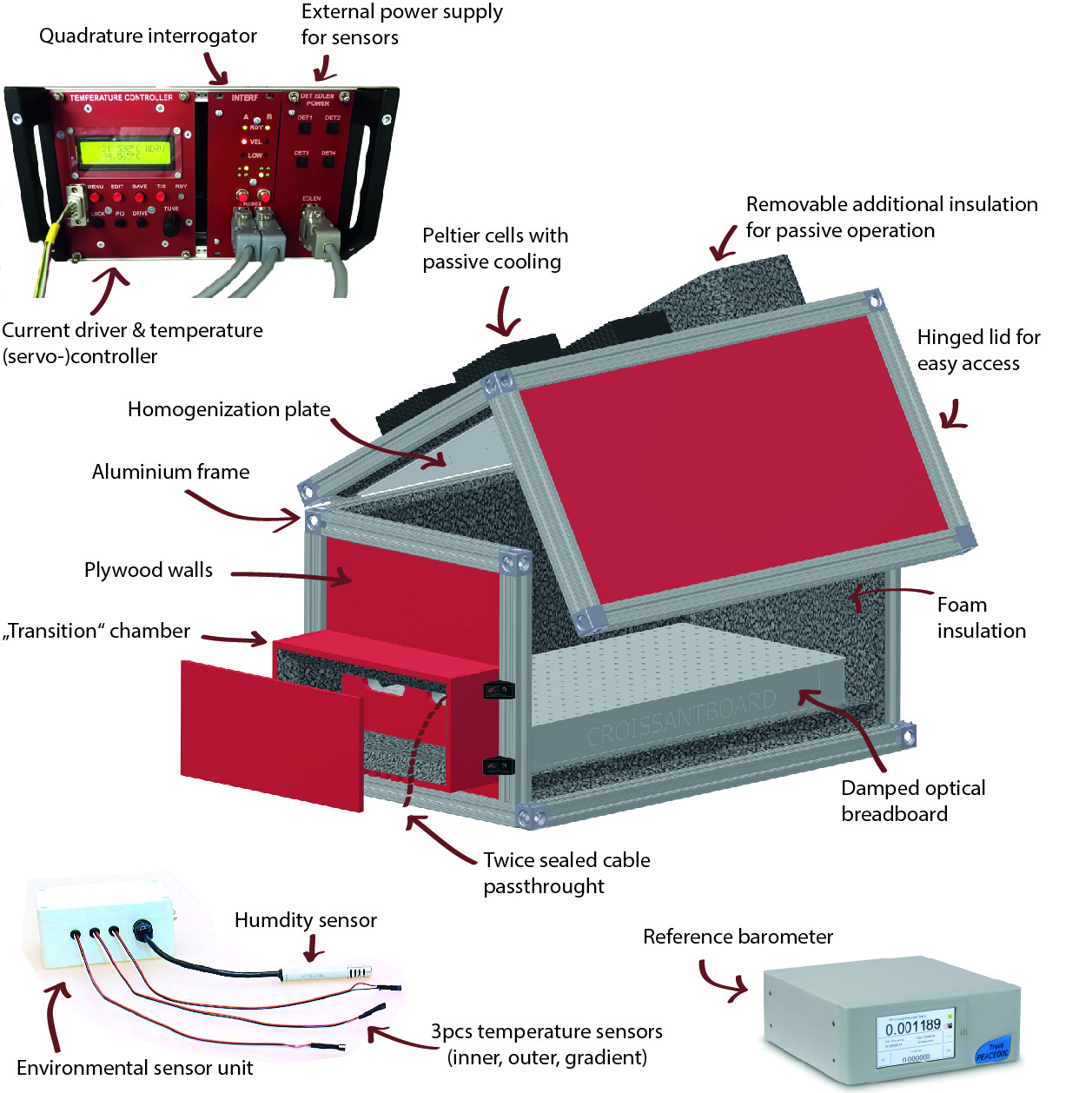} 
		\caption{The ecosystem of the temperature-controlled enclosure, including the electronics and sensorics}
		\label{fig_arrangement}
	\end{center}
\end{figure}

The entire ecosystem, sketched in Figure \ref{fig_arrangement}, consists of the actual enclosure, set of environmental sensors and control electronics.
The enclosure with a volume of $\approx 45\,$litres is designed to accomodate a 300mm x 600mm optical breadboard. The walls ($10\,$mm Birch plywood) are lined with a polyethene insulation foam ($20\,$mm plates) to provide the passive insulation. 
The enclosure's floor holds the breadboard (for versatile assembly of tested devices), and its honeycomb construction ensures mechanical stability during thermal loads. 
An array of thermo-electric cells (six pieces) are mounted on the enclosure ceiling. One side is thermally connected to passive coolers outside the box, and the other to a homogenization plate inside ($2\,$mm aluminium). 
The enclosure is designed to be operated with the temperature set slightly below the room temperature as we aim to limit the temperature difference between the chamber volume and the outside temperature and reduce the vertical temperature gradient in the chamber. This way, the warmer air rises upwards (natural convection), where the Peltier cells dissipate the heat out of the chamber, and the cooler air sinks down. 

The chamber is equipped with a range of sensors (see also Figure \ref{fig_setup}). 
Several temperature sensors monitor the temperature conditions.
The \textit{servo temperature} sensor ($T_{\mathrm{servo}}$) serves as a process variable for the servo-loop temperature control.
The \textit{reference temperature} ($T_{\mathrm{ref}}$) sensor with a calibrated absolute scale monitors the temperature at the same point as the servo-temperature sensor.
The other pair monitors the \textit{external temperature} ($T_{\mathrm{external}}$) outside the enclosure, and the \textit{gradient temperature} ($T_{\mathrm{gradient}}$) is placed above the monitor and servo sensors (with defined vertical spacing $h$) to detect the vertical temperature gradient. Besides the temperature, the air's barometric pressure and relative humidity are measured.

The control electronics comprise three integral modules.
The temperature controller TC interrogates the servo temperature (with $0,4\,$mK resolution) and drives the current to the thermo-electric cells (up to $1,9\,$A).
The interferometric modul IFM interrogates the quadrature sine-cosine output either from the homodyne optical receivers or similar interface for the case of commercial instrumentation tested in the chambers. 
The environmental sensor interrogator unit E.I.U (originally an indirect refractometer \cite{hucl13automatic}) handles the rest of the sensors: three temperature sensors (Analog Devices AD22100) with the resolution of $3\,$mK were calibrated with $U = 0.08\,$K (range from $20\,^{\circ}$C to $29\,^{\circ}$C, $k = 2$), the pressure sensor (NXP MPXH6101A; resolution $>1\,$Pa, range from $85\,$kPa to $102\,$kPa, $U = 0,01\,$kPa) and sensor for the relative humidity (Humirel HTM2500; $U < 1,5\,(\% RH)$). 
A separate module, PWR, powers the E.I.U.
Due to precision limits, a separate reference barometer (Druck PACE 1000) with $1,5\,$Pa precision was operated along the enclosure.

\subsection{Thermal conditions in the enclosure}

\begin{figure}[htbp]
	\begin{center}
		\includegraphics[width=0.8\textwidth]{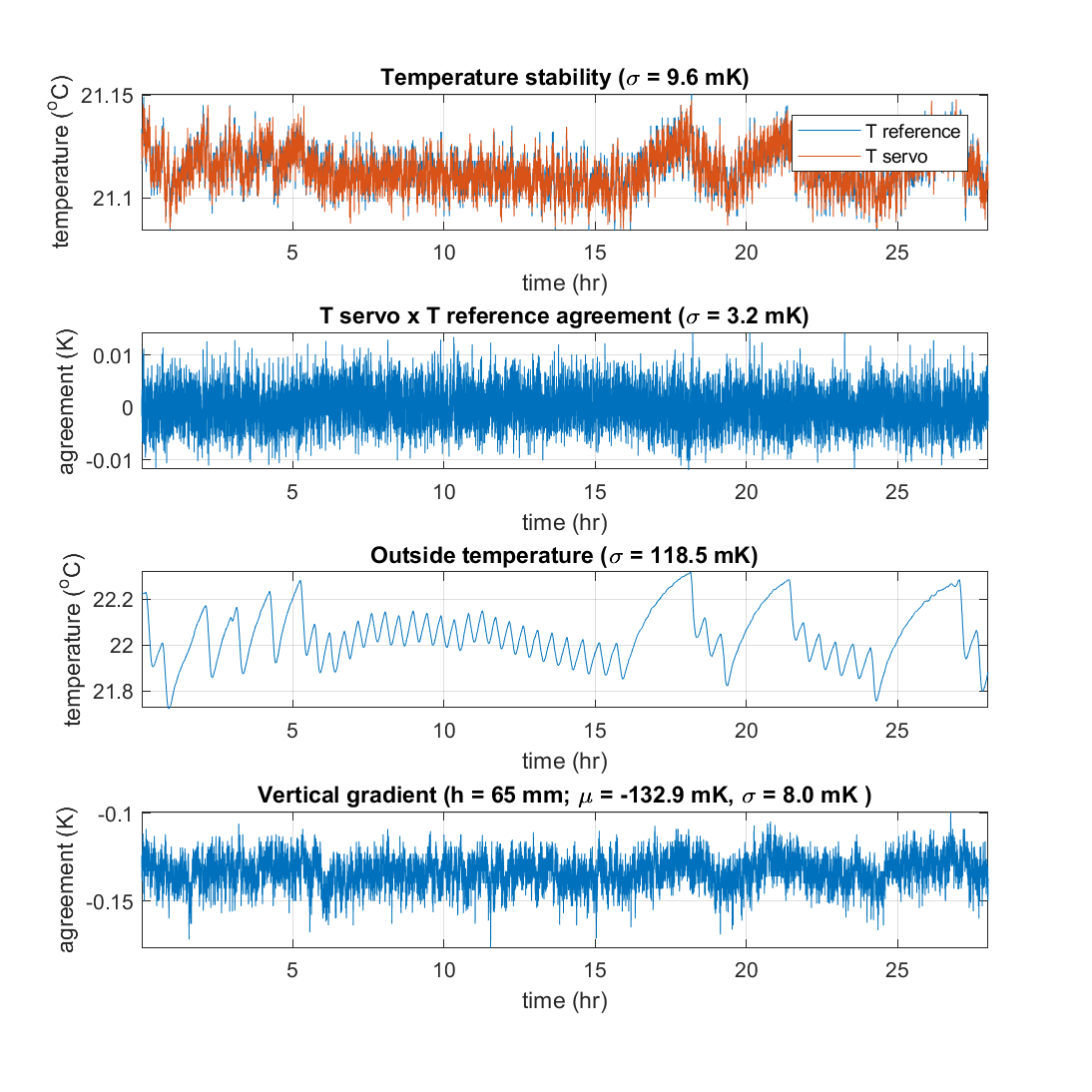} 
		\caption{Thermal conditions in the temperature-controlled enclosure: the temperature from control thermometer $T_{\mathrm{servo}}$ and reference thermometer $T_{\mathrm{ref}}$ (a) are in good agreement at the level of the $T_{\mathrm{ref}}$ resolution (b); apparently, the temperature outside the enclosure (c) have a visible influence on the temperature stability inside the enclosure and also on the vertical temperature gradient (d) present in the enclosure}
		\label{fig_thermocond01}
	\end{center}
\end{figure}

To obtain insight into the thermal conditions in the enclosure, we analyzed $\approx$day-long recording from the temperature sensors, displayed in Figure \ref{fig_thermocond01}.
The temperature fluctuated with $\sigma = 9,6\,$mK as measured by both the servo temperature sensor and the reference temperature sensor.
The good relative coincidence of the two sensors ($\sigma = 3,2\,$mK), which is at the level of resolution of the less precise sensor, proves there are no long-term drifts in the temperature measurement.
The recording of the external temperature (outside the box) indicates that the external and inner temperatures are coupled (and further tuning of the servo-control is necessary).
Finally, the measurement revealed a negative vertical temperature gradient inside the enclosure of approximately $-2,04\,$K.m$^{-1}$, which is quite significant and requires further investigation, as well as the homogeneity within the enclosure in general. Preliminary experiments carried out with a smaller version of the enclosure ($\approx18$ litre volume with $30\,$mm insulation) indicate achievable temperature stability below $\sigma = 5\,$mK over a week and the horizontal gradient below $\sigma = 1\,$K.

\subsection{Typical experimental test-bed and methodology}
\label{sec-testbed}

\begin{figure}[htbp]
	\begin{center}
		\includegraphics[width=\textwidth]{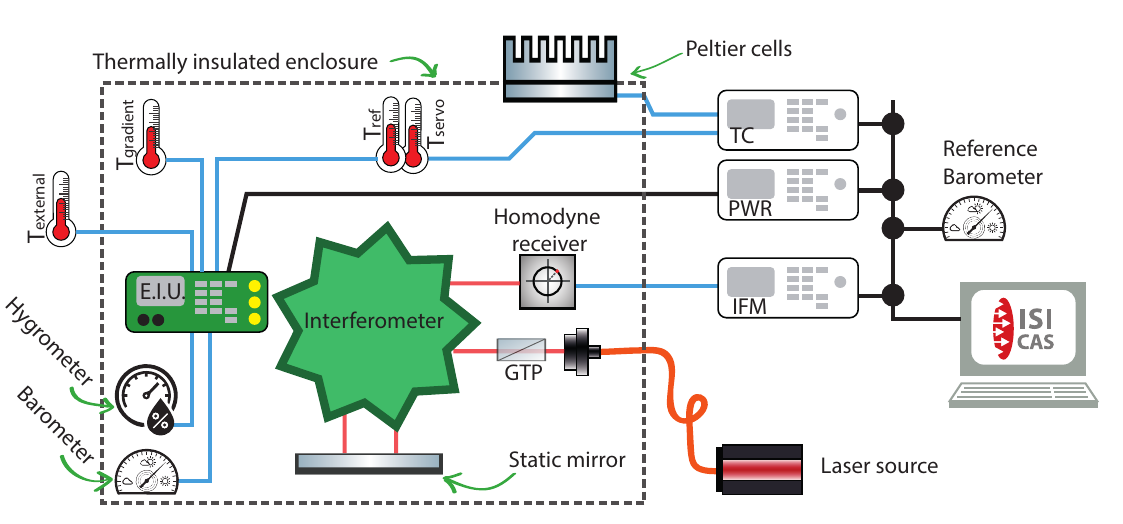} 
		\caption{The experimental assembly in a block diagram: the interferometer is usually adjusted with a single, fixed mirror inside a thermally insulated chamber; the chamber is equipped with refractometric sensors and active temperature stabilization; TC - temperature controller, GTP - calcite polarizer, E.I.U - environmental interrogation unit, PWR - power supply for E.I.U, IFM - interferometric interrogator}
		\label{fig_setup}
	\end{center}
\end{figure}

An outcome of preliminary testing (not covered in this paper) is a somehow standardized experimental arrangement and protocol for the experimentation. In a typical arrangement, depicted in Figure \ref{fig_setup}, the interferometer was adjusted against a single, fixed mirror, keeping the separation between the interferometers' front side and the mirror minimal. 

The servo-controlled temperature in the enclosure allowed for two types of investigation: (i) the temperature is kept stabilized (actively or only passively) for a prolonged period (days to weeks) so that the limits of the thermo-mechanical stability of individual components or more complex assemblies can be investigated, or (ii) the temperature was modulated (and thus the thermal load applied) in predefined cycles with linear ramp-up. 

With the bulk interferometers, the feasible temperature change rate was experimentally identified to be between $0,1\,$K$\cdot$hr$^{-1}$ and  $0,2\,$K$\cdot$hr$^{-1}$. This value represents a trade-off between the quite a small power rating of the cooling relative to the enclosure volume plus the need to let all components change the temperature as homogeneously as possible and the need to conduct the experiment in a timely manner. The preliminary experimentation revealed that higher heating/cooling rate led to more destabilized thermal conditions in the enclosure and, in turn, distorted measurement data (e.g. significant presence of overshoots in the displacement indications, apparent phase delay between the temperature modulation and displacement response, noticeable hysteresis in the correlation plot). An open question for future investigation is using harmonic temperature modulation; this approach has the potential to reduce the experimental time, but the risky aspect is that the thermal response of the investigated system might get attenuated with higher modulation frequencies and that the attenuation might not be as apparent as it is now with the trapezoidal modulation profile. 

The temperature sensors were placed on a pillar post close to the center of the enclosure, with the servo and reference sensor being but to the same height as the interferometers' optics and the gradient sensor above them ($h = 65\,$mm). 

The laser beam was delivered using a polarization-maintaining fibre either from a frequency-stabilized He-Ne laser or from an iodine-stabilized semiconductor laser system \cite{rerucha2017dbr633}. 
Any residual fluctuation in the incoming beam polarization was filtered out with a calcite polarizer (e.g. Glan-Thompson type).
When suitable, the interferograms were processed with a four-detector type homodyne receiver (shown in Figure \ref{fig_ifms} d) and described in \cite{rerucha2012detection}). The symmetric nature of this receiver architecture provides good resilience against amplitude fluctuations of the interferogram signal.
The interferometer-mirror-receiver triplet(s) were adjusted with the help of a digital oscilloscope to obtain a symmetric and well-centred Lissajeous diagram at the receiver's output. 
The analogue output of the receiver is digitized with the IFM module. Any residual offset in the analogue-to-digital conversion is calibrated before the measurement, and the drifts of the offset are negligible.

\begin{figure}[htbp]
	\begin{center}
    \includegraphics[width=\textwidth]{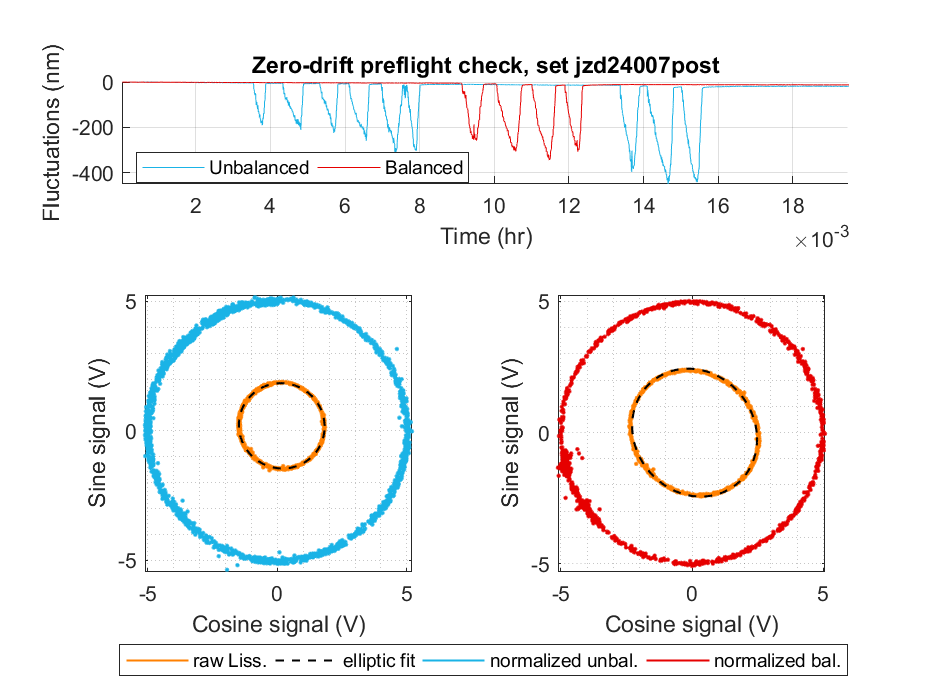}
    \caption{The interferometers were checked before the test run for correct counting direction, and data for elliptic correction were collected}
	\label{fig_preflight}
	\end{center}
\end{figure}

The data from the homodyne receivers, control modules and sensors was recorded using a unified communication framework (CANopen protocol with custom extension) at a native sampling rate of individual modules (between $5\,$Hz and $500\,$Hz), subsampled to $2\,$Hz and evaluated. Before the actual test run, after the interferometers were adjusted, we manually induced vibrations to the system and simultaneously recorded the response to (i) double-check the counting direction and (ii) obtain data for elliptic correction \cite{cip2000scale}, which was later applied to the recorded data to remove any remaining offsets or asymmetry in the quadrature output; example recording is shown in Figure \ref{fig_preflight}. 
The follow-up data analysis incorporated the removal of a residual linear drift, applying the linearity correction and the refractive index of air compensation, evaluating the correlation between the induced temperature modulation and drifts in the interferometers' displacement reading, and final interpretation.

The total relative uncertainty associated with the refractive index measurement was $U = 8.5\times10^{-8}$ ($k=2$), so with the distance between the front face of the interferometer and the mirror $<3\,$mm, the absolute uncertainty was $<1\,$nm. A significantly larger contribution could be expected from the thermal expansion of the experimental optomechanics, as discussed later.

\section{Experimental investigations with the enclosure}
\label{s:experimental}

In the following sections, we present particular experimental studies exploiting the (thermally-induced) zero-drift investigation to characterize or verify displacement-measuring interferometers. 

The first experimental measurement investigates the (primarily temperature-induced) zero drift in two variants of single-ended interferometer arrangement and compares the result to the theoretically estimated figures. This experiment, described in section \ref{sec-test1}, demonstrates how to characterize the zero-drifts and thus, e.g. establish the upper boundaries for uncertainty assessment for interferometric measurements or allow for deterministic compensation of the effects.
This is significant both in less-controlled environments and in longer time-frames -- in both cases, both the temperature-induced and general zero-drift where the temperature effects are of more substantial importance. 

Once the zero-drift behaviour is characterized for a particular interferometer, this knowledge is useful as a basis for (quality) testing or verification. This aspect is demonstrated in the second experimental study (section \ref{sec-test2}), where the experimental test revealed an assembly failure in the interferometer's optics that would be otherwise hardly detectable.

Also, knowing the zero-drift of the interferometer's optics, it is possible to extend the characterization to a more complex assembly. In section \ref{sec-test3}, we present how the thermal-load testing provided useful feedback in adjusting the mechanical fixture, i.e. it helped verify the system's thermal stability.

Finally, thermal testing can help aid the design or manufacturing decision; in the case of interferometric optics, it is, for example, the assembly technology. In section \ref{sec-test4}, we present the investigation and comparison of three interferometers of identical geometry and dimension but with different assembly methods and compare the experimental results with theoretical predictions.

All the use cases represent several experimental scenarios that address different research questions revolving around studying the thermal drifts in length measurement instrumentation. Combined, they give an indicative overview of the range of testing and characterization possibilities.

\subsection{Comparing variants of interferometric arrangement}
\label{sec-test1}

\begin{figure}[htbp]
	\begin{center}
		\includegraphics[width=\textwidth]{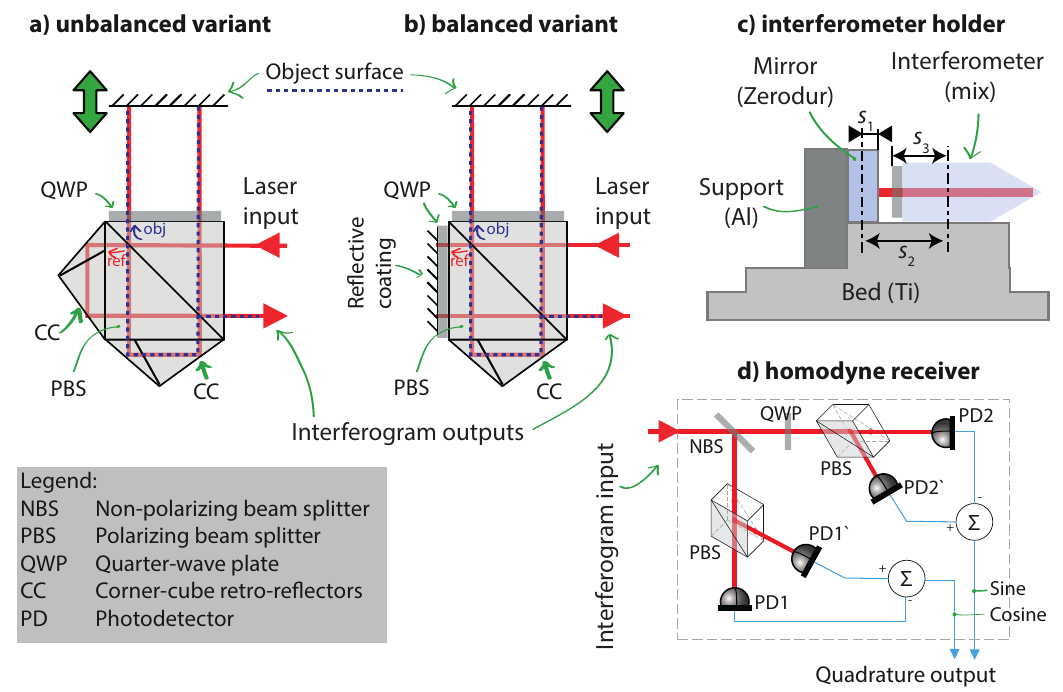} 
		\caption{The optomechanics used for the thermal sensitivity comparison of planar interferometers: the conventional version with unbalanced pathlengths (a), the modified variant with balanced paths (b), the holder from low-expansion material (c), the four-detector homodyne receiver with quadrature output (d)}
		\label{fig_ifms}
	\end{center}
\end{figure}

In this experiment, we have compared two similar two-beam double pass single-ended (i.e. non-differential) planar interferometers, displayed in Figure \ref{fig_ifms} a) and b). This type of interferometer is commonly used, e.g. for coordinate positioning. The two variants have slightly different layouts of the optical paths of the two arms, and, as a result, the optical path difference within the interferometers' optics is significantly different.

In the variant with two cube-corners (Figure \ref{fig_ifms} a)), the reference beam passes through the central beam-splitter twice, while the object beam does four passes and, additionally, it goes through the quarter-wave retarder (QWP) four times. Thus, the optical paths will differ in length by OPD $= 2a + 4b$, where $a$ is the nominal dimension of the beam-splitter and $b$ stands for the waveplate thickness. 
In contrast, in the other arrangement (Figure \ref{fig_ifms} b)), both arms pass the beam-splitter and either of the QWPs four times, so the beam paths within the optics are balanced.

Based on this consideration, we can estimate the optic thermal drift in the \textit{unbalanced} interferometer (the drift presence of which has been mentioned decades ago \cite{steinmetz1990sub}). The interferometers's beam splitters were made of N-SF14 flint glass (Schott) and their nominal dimension was $15\,$mm. The waveplates of crystal quartz had a thickness of $0,4\,$mm. Thus, the beam paths differ by $31,6\,$mm, so the optical path difference (incorporating thermal expansion and thermo-optic coefficient) would drift by $241,7\,$nm per kelvin, so the total thermal sensitivity of the displacement measurement will be $60,43\,$nm.K$^{-1}$ (considering the double-pass layout). With the balanced arrangement, the thermal drifts are expected to reduce to zero. 

\begin{figure}[htbp]
	\begin{center}
		\includegraphics[width=\textwidth]{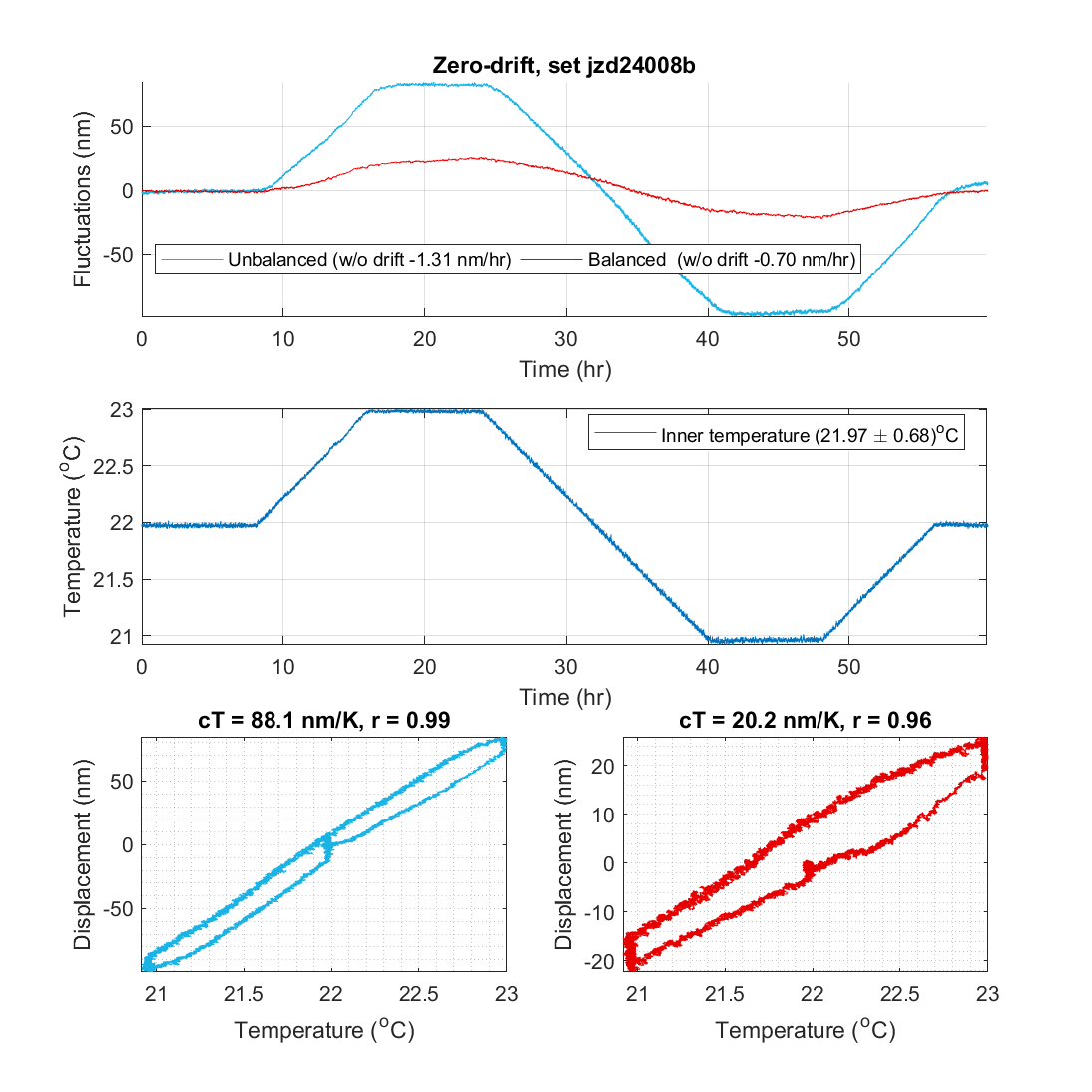} 
		\caption{Comparison of temperature-induced length drifts in two similar interferometers: interferometer with unequal length of beam paths in the individual arms exhibits significantly higher drift (a) under thermal load (b); the drifts are highly correlated with the temperature modulation (c) }
		\label{fig_data1}
	\end{center}
\end{figure}

We have recently carried out a similar study with two variants of a differential interferometer\cite{rerucha2021compact} and demonstrated significant improvement in the thermal resilience \cite{rerucha2024interferometer}. The study presented here, particularly the quantitative assessment of the thermal sensitivity, is significantly complicated because the interferometers are not of a differential arrangement. Consequently, the metrological loop is considerably longer and more complex, e.g. the physical distance between the interferometer is also susceptible to thermal expansion and adds up to the observed drifts. 

In this experiment, the interferometers were mounted on dedicated holders made of low expansion Titanium (depicted in Figure \ref{fig_ifms} c)) and aligned against planar mirrors (Zerodur + dielectric coating). While it is a straightforward choice to use the low-expansion materials for the optomechanics, there is also a significant drawback: the same materials also exhibit low thermal conductivity, so in turn, it poses a further limit on the temperature loading bandwidth.

To estimate the contribution of the mounting, we assume both the mirror and the interferometer to be fixed to the holder (bed) at the centre of gravity (marked with the dash-dotted lines). Then we accounted for the relevant material amount of the mirror (denoted $s_1$, $s_1 = 2,5\,$mm), the bed ($s_2 = 11,4\,$mm) and the interferometer ($s_3 = 7,9\,$mm). The total contribution to the temperature sensitivity was approximately $24,67\,$nm.K$^{-1}$. Nonetheless, the contributions of individual terms $s_{1,2,3}$ reaches almost a hundred nanometers, so we expect quite high uncertainty associated with this estimation.

For this experiment, we used the temperature change of $\pm1\,$K with the cooling/heating rate of $0,12\,$K/hour. 
The experimental results are shown in Figure \ref{fig_data1}.
The interferometers' displacement reading have a residual linear drift subtracted: $1,31\,$nm.hr$^{-1}$ for the unbalanced interferometer and $0,67\,$nm.hr$^{-1}$ for the balanced one. 
The observed drifts were highly correlated to the induced thermal cycling ($r >= 0,96$), but the observed hysteresis indicates the presence of either a residual drift component or too steep temperature change. 

Subtracting the contribution of the holder, the total observed thermal sensitivity is $63,43\,$nm.K$^{-1}$ for the unbalanced interferometer and $-4,47\,$nm.hr$^{-1}$ for the balanced one. These figures correspond well to the estimations (coindcidence $<5\,$nm.K$^{-1}$). Nonetheless, we expect the associated uncertainty to be considerable (probably $10\,$nm.K$^{-1}, k = 2$ or more) for a number of reasons, where the imperfection of the mounting thermal response model (with several questionable parameter values, such as the effective mounting pivot points) would be the dominant one. Also, the glue layers between the optical elements of the interferometer and the dielectric coatings remain unaccounted for. On the other hand, with the small beam path on the air ($>\approx1\,$mm), the influence of the refractive index of air would be less than $1\,$nm.K$^{-1}$ combined.

\subsection{Identifying faulty optical assembly}
\label{sec-test2}

\begin{figure}[htbp]
	\begin{center}
		\includegraphics[width=\textwidth]{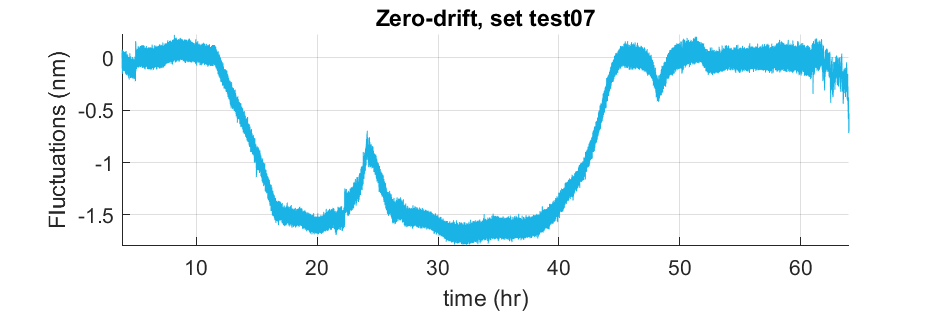} 
		\includegraphics[width=\textwidth]{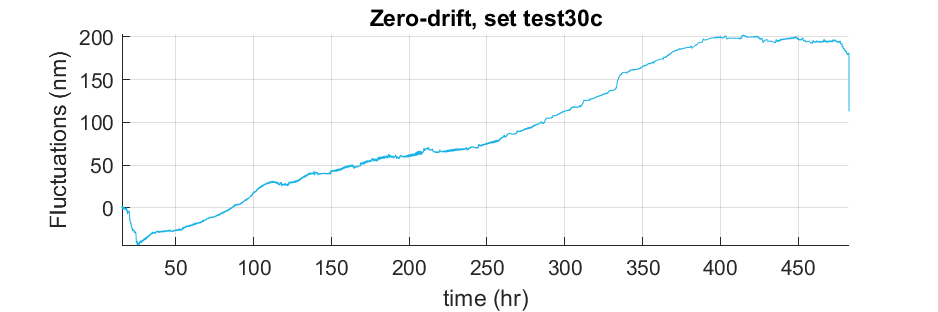} 
		\caption{Comparison of thermal drifts in two differential interferometers of identical design\cite{rerucha2024interferometer}: while both pieces appeared perfectly functional, in comparison to the reference results (top) the tested sample exhibited a significant long-term zero drift (bottom); a hardly noticeable assembly fault was later identified in the optical assembly)}
		\label{fig_data2}
	\end{center}
\end{figure}

The aim of the test was to verify that a manufactured sample of a double-pass planar differential interferometer \cite{rerucha2024interferometer}. 
We tested whether it has the expected thermal sensitivity compared to the previously observed performance we achieved with the reference interferometer of the same construction (see the top plot in Figure \ref{fig_data2}). 
With this interferometer, we observed the drifts in the range smaller than $2\,$nm peak-to-peak over $>60\,$ hour timeframe with the temperature fluctuation of $\sigma = 0,19\,$K (the measurement did not take place in the enclosure). The residual fluctuations were attributed to the polarization fluctuations in the fibre feed.

The interferometer under test was put in the enclosure (without the active thermal control at that time) to investigate the drifts using the standardized experimental testbed and protocol (Figure \ref{fig_setup}).  The interferometer was mounted flat onto a raised platform with a $2"$ silver mirror in kinematic mount, placed $\approx 2\,$mm apart from the interferometer's front side. 

The tested interferometer did not show any signs of malfunctions (such as poor contrast, increased noise, presence of etalon effects, or quadrature signal distortion). 
The test in the enclosure revealed significant drifts (see the lower part in Figure \ref{fig_data2}) of hundreds of nanometres (over several weeks; achieved temperature stability was $\sigma = 0,15\,$K) instead of expected fluctuation at a nanometre level.
After all other components were gradually replaced in the measurement chain (laser, fibre feed, receiver, mirror with holder) to rule out their contribution, the interferometer underwent partial disassembly. 
A loosened single corner of a retarder waveplate was revealed and identified as the source of drifts.

The moral of this experiment is that zero-drift testing has a significant potential not only in characterizing the drifts but also in verifying the conformance of the interferometers (and generally any measuring instruments). 
The prior knowledge that is obviously necessary could come either from the previous experimental results or from analytical sources (simulations, calculations). 
The two significant benefits are that the technique could reveal faults otherwise difficult to detect and that the approach is simplistic enough e.g. to be useful for 100\% verification of optical systems during manufacturing. 

\subsection{Improving mounting stability}
\label{sec-test3}

Another measurement scenario demonstrates that the range of testing possibilities with the enclosure reaches beyond the interferometer optics itself. 
When asked to characterize a pair of commercial interferometers, we used the standard experimental arrangement to test both interferometers simultaneously. 
The interferometers were mounted vertically according to the mechanical interface specified in the device documentation. Generic opto-mechanical components were used as the baseplates.

\begin{figure}[htbp]
	\begin{center}
		\includegraphics[width=\textwidth]{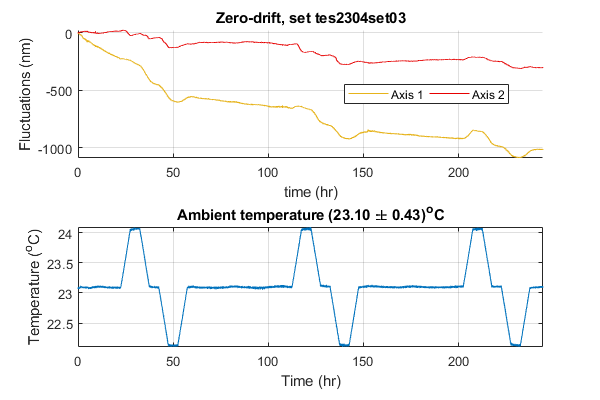} 
		\includegraphics[width=\textwidth]{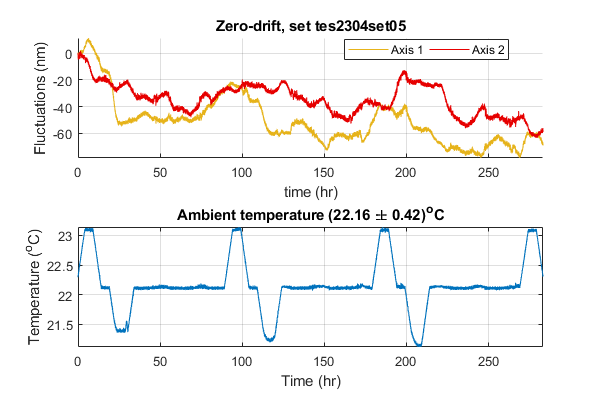} 
		\caption{The influence of the mechanical fixture on the observed zero-drifts in commercial differential interferometers: the straightforward fixture, according to manufacturer documentation, led to a significant global drift (upper part); the fixture reinforced with additional supports reduced the drift significantly (lower part)}
		\label{fig_data3}
	\end{center}
\end{figure} 

With the rough idea (the manufacturer did not disclose the detailed arrangement) of the differential double-pass interferometer's arrangement, we expected fairly pronounced thermal drift. 
The test results, however, revealed a significant global drift of several hundred nanometres (see the upper part in Figure \ref{fig_data3}).  
In several consecutive attempts, the fixture was gradually replaced with more robust components with enhanced triangular supports and additional side supports were added.
With these measures applied, the observed drift was reduced by a factor of five with one of the interferometers and even more with the second one of the pair (see the lower part in Figure \ref{fig_data3}). Notably, the effect of the thermal load was still less significant (which was a valuable outcome of the test) than the residual (global) drift.

With the analysis of thermal influences presented before in mind, this type of test could provide a valuable tool in the characterization and verification of complex (sub-) systems intended to become part of measurement instrumentation.
It is also evident that the individual manifestations of thermal changes are principally superimposed and challenging to discern, and their mitigation and fine-tuning requires careful consideration in multiple iterations. The zero-drift analysis can aid the effort and provide necessary feedback during the process.

\subsection{Comparison of the assembly technology}
\label{sec-test4}

\begin{figure}[htbp]
	\begin{center}
		\includegraphics[width=\textwidth]{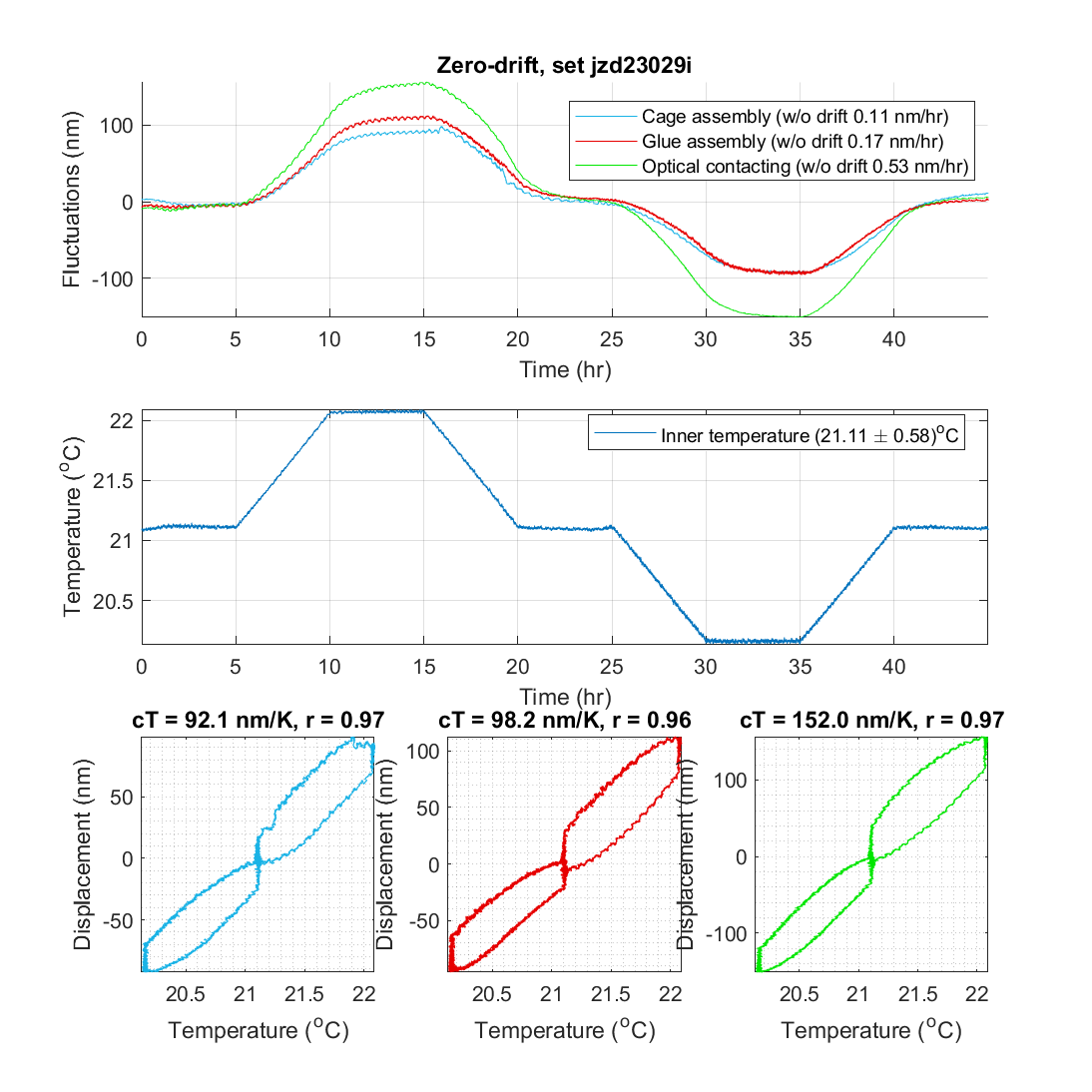} 
		\caption{Comparison of the thermally-induced drifts of three interferometers with  different assembly processes: the observed drifts (top), the temperature during the test (middle) and the correlation between the temperature inside the enclosure and the drift}
		\label{fig_data4}
	\end{center}
\end{figure} 

The last experimental investigation we present is similar to comparing interferometers with the slightly varied arrangements (Section \ref{sec-test1}).
This time, we loaded the enclosure with three interferometers of identical arrangement (the unbalanced planar interferometer, depicted in Figure \ref{fig_ifms} a)) but differing in the assembly technique and, consequently, the glass material of the bulk components. The goal was to compare the thermal sensitivity to otherwise identical interferometers experimentally. 

The first interferometer has the components separately glued into the metal (aluminium) housing (cage assembly). In the second case, the optical components were cemented together (glue assembly), and the last interferometer was assembled using optical contacting (wringing). 
The first two interferometers were made of the N-BK7 and N-SF14 glasses (Schott), while the last one was made of fused silica. As mentioned before, the estimated thermal sensitivity of the N-BK7+N-SF14 interferometers were estimated to be $\approx60$nm.K$^{-1}$. Fused silica has a lower coefficient of thermal expansion, nonetheless, it has a significantly higher thermo-optic coefficient, so the temperature influence on the optical pathlengths more pronounced, and the total thermal sensitivity is $\approx100$nm.K$^{-1}$.

The interferometers were mounted on a titanium holder (Figure \ref{fig_ifms} c)) with a $1"$ silver mirrors (instead of the rectangular Zerodur mirrors used in Section \ref{sec-test1}). This change and using a larger separation between the mirror and the front face of the interferometer caused the contribution to the zero drift to rise to approximately $40\,$nm.K$^{-1}$.

The results of the test, displayed in Figure \ref{fig_data4}, indicate the total temperature sensitivity of the assembly (interferometer + mirror + holder) is between $90\,$nm.K${^-1}$ and $145\,$nm.K${^-1}$. 
Similarly to the experiment from Section \ref{sec-test1}, the observed drifts were well correlated to the induced thermal cycling ($r >= 0,96$), and noticeable hysteresis of (a symmetric nature) indicating too steep temperature change was observed. 

Taking the contribution of the holder, we get approximate coefficients as follows: cage assembly $51\,$nm.K${^-1}$, glue assembly $56\,$nm.K${^-1}$, optical contacting $104\,$nm.K${^-1}$. All figures coincide with the theoretical estimation within $10\,$nm.K${^-1}$. With the cage assembly, the physical separation of the components (unlike the other two assemblies) probably makes the interferometer partially self-compensated against the thermal drifts.

\section{Final words}
\label{s:final}

We have presented the achieved status in designing and developing a thermal-controlled enclosure dedicated to testing and characterizing displacement-measuring interferometers and their assemblies. The servo-loop temperature control achieves temperature stability of $\sigma = 9,6\,$mK during a day and enables setting the temperature in the range of several kelvins. 
The enclosure could be used to keep the temperature stable over days and weeks, and it could also be used to apply a controlled thermal load to the devices under test. 

We demonstrated four use cases that used either of the capabilities for various characterization or testing tasks. 
The comparisons of zero-drift between similar interferometers (different assembly techniques,  variant optical arrangement) provided valuable feedback for expanding the capabilities of and refining the interferometric measurement technology. Generally, this type of study can aid the development process, provide experimental verification of the instrumentation design, and provide critical input to the uncertainty assessment. 
Similarly, using the thermal loading for intermittent testing provided feedback for redesigning the interferometer's mechanical fixtures, significantly reducing the drifts and demonstrating the feasibility of using the zero drift investigations for iterative improvement and verification in the development of metrological instrumentation.
Similarly, the methodology proved feasible for routine verification of the optical and optomechanical systems, where the testing under thermal load could help identify otherwise undetectable failures in the systems. 

The results from the presented investigations also have a potential impact on particular applications. The observed magnitude of thermally-induced zero drifts might interfere with the accuracy requirements in the long-range coordinate positioning ($100+\,$mm), either in air (nanometrology) or in vacuum (precise microscopy, e-beam lithography), also in 1-D length calibration instrumentation for secondary- or shop-floor metrology in the range up to $1\,$m (the influence of the refractive index would prevail above this level).

Last but not least, the construction of the temperature-controlled enclosure and characterization of the thermal conditions inside it could be transferred to the design of measurement instrumentation:
Within the future development of dimensional metrology instrumentation, proper, intelligent and highly precise thermal management should become an intrinsic part of the design (e.g. \cite{hamid2005temperature}) not only in the top-grade instruments at national metrology laboratories -- this an appealing approach to carrying out measurements under defined conditions, and simultaneously a necessary condition on the way to lower measurement uncertainty and an extended range of measurement scenarios.

While the particular quantitative results we presented might be considered indicative rather than absolute, the message and contribution of this paper to our ultimate belief are demonstrative and methodological. The current state of the work opens both apparent questions and challenges as well as the broader questions with overlap to precision engineering and measurement technology (e.g. investigation of the temperature distribution in the thermally controlled environment, because the temperature stability still relies on a single point measurement). 
Back to the interferometry, we also believe this work is crucial on the way towards further expansion of interferometric technology to routine operation in longer timeframes or in a less controlled environment -- for instance, in the heart of a nuclear power plant\cite{rerucha18temelin}.

\section*{Acknowledgement}
Technology Agency of the Czech Republic (FW03010687;TN02000020); 
European Commission (CZ.1.05/2.1.00/01.0017, LO1212, CZ.02.1.01/0.0/0.0/16\_026/0008460,  CZ.02.01.01/00/22\_008/0004649);
Czech Academy of Sciences (RVO:68081731)

\section*{Disclosure}
The authors declare no conflicts of interest. While preparing this work, the authors used the Grammarly tool (www.grammarly.com) to correct the language. After using this tool/service, the authors reviewed and edited the content as needed and take full responsibility for the content of the published article.

\section*{Data Availability Statement}
The authors are willing to share the data that support the findings of this study on the basis of a reasonable request.

\section*{Authors' Contribution}

\textit{S. Rerucha}--  Conceptualization, Funding acquisition, Investigation, Methodology, Software, Data Curation, Validation, Visualization, Writing - Original Draft, Writing - Review and Editing.
\textit{M. Hola} -- Investigation, Resources (optics, experimental setups).
\textit{J. Oulehla} -- Resources (optical coatings), Investigation, Data Curation.
\textit{J. Lazar} -- Conceptualization, Methodology.
\textit{B. Mikel} -- Project management, Supervision.
\textit{O. Cip} -- Conceptualization, Funding acquisition, Supervision.
\textit{All authors} -- Writing - Review and Editing.


\section*{References}
\bibliographystyle{unsrt}
\bibliography{references}

\begin{thebibliography}{10}

\bibitem{yang2018review}
S.~Yang and G.~Zhang.
\newblock A review of interferometry for geometric measurement.
\newblock {\em Measurement Science and Technology}, 29(10):102001, 2018.

\bibitem{michelson1887ontherelative}
Albert~A Michelson and Edward~W Morley.
\newblock Lviii. on the relative motion of the earth and the luminiferous
  {\ae}ther.
\newblock {\em The London, Edinburgh, and Dublin Philosophical Magazine and
  Journal of Science}, 24(151):449--463, 1887.

\bibitem{coveney2020review}
T.~Coveney.
\newblock A review of state-of-the-art {1D} length scale calibration
  instruments.
\newblock {\em Measurement Science and Technology}, 31(4):042002, 2020.

\bibitem{stone2014testcalibration}
J.~Stone.
\newblock Test and calibration of displacement measuring interferometers.
\newblock In {\em Proceedings of the 3rd International Conference on Mechanical
  Metrology (CIMMEC III), Gramado, Brazil.}, 2014.

\bibitem{figueroa2019ability}
D.~G. Figueroa and E.~H. Tanin.
\newblock Ability of {LIGO} and {LISA} to probe the equation of state of the
  early universe.
\newblock {\em Journal of Cosmology and Astroparticle Physics}, 2019(08):011,
  2019.

\bibitem{robinson2016watt}
I.~A. Robinson and S.~Schlamminger.
\newblock The watt or {Kib\-ble} balance: a technique for implementing the new
  {SI} definition of the unit of mass.
\newblock {\em Metrologia}, 53(5):A46, 2016.

\bibitem{rerucha2021ctrp}
S.~Rerucha, M.~Hola, M.~Sarbort, J.~Kur, P.~Konecny, J.~Lazar, and O.~Cip.
\newblock Laser-interferometric nanometre comparator for length gauge
  calibration in advanced manufacturing.
\newblock In {\em 2021 International Conference on Electrical, Computer,
  Communications and Mechatronics Engineering (ICECCME)}, pages 1--5. IEEE,
  2021.

\bibitem{rerucha18temelin}
Š. Řeřucha, B.~Mikel, Z.~Matěj, O.~Herman, M.~Holá, M.~Jelínek,
  P.~Jedlička, O.~Číp, and J.~Lazar.
\newblock {Linearized and Compensated Interferometric System for High-Velocity
  Traceable Length Calibration on a Metre Scale}.
\newblock {\em Proc. SPIE}, {10976}:39--45, {2018}.

\bibitem{rerucha2024interferometer}
Simon Rerucha, Miroslava Hola, Jindrich Oulehla, Josef Lazar, Bretislav Mikel,
  and Ondrej Cip.
\newblock Thermally compensated common-path differential interferometer with
  reduced long-term zero-drifts.
\newblock {\em Measurement Science and Technology}, 35(9):095021, jun 2024.

\bibitem{byman2018high}
V.~Byman, T.~Jaakkola, I.~Palosuo, and A.~Lassila.
\newblock High accuracy step gauge interferometer.
\newblock {\em Measurement Science and Technology}, 29(5):054003, 2018.

\bibitem{lassila2011design}
A~Lassila, M~Kari, H~Koivula, U~Koivula, J~Kortstr{\"o}m, E~Leinonen,
  J~Manninen, J~Manssila, T~Mansten, T~Meril{\"a}inen, et~al.
\newblock Design and performance of an advanced metrology building for mikes.
\newblock {\em Measurement}, 44(2):399--425, 2011.

\bibitem{manske2012recent}
E.~Manske, G.~J{\"a}ger, T.~Hausotte, and R.~F{\"u}{\ss}l.
\newblock Recent developments and challenges of nanopositioning and
  nanomeasuring technology.
\newblock {\em Measurement Science and Technology}, 23(7):074001, 2012.

\bibitem{yacoot2000use}
A.~Yacoot and M.~J. Downs.
\newblock The use of x-ray interferometry to investigate the linearity of the
  npl differential plane mirror optical interferometer.
\newblock {\em Measurement Science and Technology}, 11(8):1126, 2000.

\bibitem{rerucha2021compact}
{\v{S}}.~{\v{R}}e{\v{r}}ucha, M.~Hol{\'a}, M.~{\v{S}}arbort, J.~Hrabina,
  J.~Oulehla, O.~{\v{C}}{\'\i}p, and J.~Lazar.
\newblock Compact differential plane interferometer with in-axis mirror tilt
  detection.
\newblock {\em Optics and Lasers in Engineering}, 141:106568, 2021.

\bibitem{yague2021scalability}
Jose~A Yag{\"u}e-Fabra, Wei Gao, Andreas Archenti, Edward Morse, and Alkan
  Donmez.
\newblock Scalability of precision design principles for machines and
  instruments.
\newblock {\em CIRP annals}, 70(2):659--680, 2021.

\bibitem{weichert2018vacuumsetup}
Ch. Weichert, S.~Quabis, and J.~Flügge.
\newblock A new vacuum setup for fundamental investigations on interferometric
  length measurements.
\newblock In {\em euspen’s 18th International Conference \& Exhibition,
  Venice, IT}, 2018.

\bibitem{schodel2021new}
Rene Sch{\"o}del, Andrew Yacoot, and Andrew Lewis.
\newblock The new mise en pratique for the metre—a review of approaches for
  the practical realization of traceable length metrology from 10- 11 m to 1013
  m.
\newblock {\em Metrologia}, 58(5):052002, 2021.

\bibitem{ciddor1996refractive}
P~E Ciddor.
\newblock Refractive index of air: new equations for the visible and near
  infrared.
\newblock {\em Applied optics}, 35(9):1566--1573, 1996.

\bibitem{ghosh1997sellmeier}
Gorachand Ghosh.
\newblock Sellmeier coefficients and dispersion of thermo-optic coefficients
  for some optical glasses.
\newblock {\em Applied optics}, 36(7):1540--1546, 1997.

\bibitem{palik1997thermo}
Edward~D. Palik.
\newblock Chapter 3 - thermo-optic coefficients.
\newblock In Edward~D. Palik, editor, {\em Handbook of Optical Constants of
  Solids}, pages 115--261. Academic Press, Burlington, 1997.

\bibitem{steinmetz1990sub}
CR~Steinmetz.
\newblock Sub-micron position measurement and control on precision machine
  tools with laser interferometry.
\newblock {\em Precision engineering}, 12(1):12--24, 1990.

\bibitem{pisani2012comparison}
Marco Pisani, Andrew Yacoot, Petr Balling, Nicola Bancone, Cengiz Birlikseven,
  Mehmet Celik, Jens Fl{\"u}gge, Ramiz Hamid, Paul K{\"o}chert, Petr Kren,
  et~al.
\newblock Comparison of the performance of the next generation of optical
  interferometers.
\newblock {\em Metrologia}, 49(4):455, 2012.

\bibitem{hucl13automatic}
V.~Hucl, M.~Cizek, J.~Hrabina, B.~Mikel, S.~Rerucha, Z.~Buchta, P.~Jedlicka,
  A.~Lesundak, J.~Oulehla, L.~Mrna, M.~Sarbort, R.~Smid, J.~Lazar, and O.~Cip.
\newblock Automatic unit for measuring refractive index of air based on ciddor
  equation and its verification using direct interferometric measurement
  method.
\newblock {\em Proc. SPIE}, 8788, 2013.

\bibitem{rerucha2017dbr633}
Simon Rerucha, Andrew Yacoot, Tuan~M Pham, Martin Cizek, Vaclav Hucl, Josef
  Lazar, and Ondrej Cip.
\newblock Laser source for dimensional metrology: investigation of an iodine
  stabilized system based on narrow linewidth 633 nm dbr diode.
\newblock {\em Measurement Science and Technology}, 28(4):045204, 2017.

\bibitem{rerucha2012detection}
Simon Rerucha, Zdenek Buchta, Martin Sarbort, Josef Lazar, and Ondrej Cip.
\newblock Detection of interference phase by digital computation of quadrature
  signals in homodyne laser interferometry.
\newblock {\em Sensors}, 12(10):14095--14112, 2012.

\bibitem{cip2000scale}
Ondrej Cip and Frantisek Petru.
\newblock A scale-linearization method for precise laser interferometry.
\newblock {\em Measurement Science and Technology}, 11(2):133, 2000.

\bibitem{hamid2005temperature}
R~Hamid, D~Sendogdu, and C~Erdogan.
\newblock The temperature stabilization and temperature measurement of a
  k{\"o}sters interferometer.
\newblock {\em Measurement Science and Technology}, 16(11):2201, 2005.

\end{thebibliography}

\end{document}